\documentclass[12pt]{article}

\usepackage{amsmath}
\usepackage{amsthm}
\usepackage{amssymb}
\usepackage{mathrsfs}
\usepackage{bbm}
\usepackage{graphicx,color}

\newcommand{\cf}{{\it cf.\/} }

\newcommand{\ie}{{\it i.e.\/} }
\newcommand{\eg}{{\it e.g.\/} }

\newcommand{\iu}{\mathrm{i}}
\newcommand{\Id}{1}
\newcommand{\alg}[1]{\mathcal{A}(#1)}

\newcommand{\str}{^{*}}
\newcommand{\ep}[1]{\mathrm{e}^{#1}}
\newcommand{\hilb}{\mathcal{H}}
\newcommand{\dd}{\mathrm{d}}
\newcommand{\tr}{\mathrm{tr}}
\newcommand{\Tr}{\mathrm{Tr}}
\newcommand{\mydet}{\mathrm{det}}

\newcommand{\mean}[1]{\langle #1 \rangle}
\newcommand{\cumul}[1]{\langle\!\langle #1 \rangle\!\rangle}

\title{Charge transport and determinants}

\begin{document}
\author{S. Bachmann and G.M. Graf\\
\normalsize\it Theoretische Physik, ETH-H\"onggerberg, 8093 Z\"urich, Switzerland \\}
\maketitle

\begin{abstract} We review some known facts in the transport theory of
mesoscopic systems, including counting statistics, and discuss its relation 
with the mathematical treatment of open systems. 
\end{abstract}


\section{Introduction}


The aim of these notes is to introduce to some theoretical developments
concerning transport in mesoscopic systems. More specifically, we intend to
show how concepts and tools from mathematical physics provide ways and means
to put some recent, fundamental results on counting statistics on rigorous
ground and in a natural setting. We will draw on concepts like C*-algebras,
which have been often used in the mathematical treatment of systems out of
equilibrium, see \eg\cite{JP}, but also on tools like Fredholm determinants, which have been used for renormalization purposes in quantum field theory. Before going into mathematical details we will review some of the more familiar aspects of transport, and notably noise. That will provide some examples on which to later illustrate the theory.

These notes are not intended for the expert. On the contrary, the style 
might be overly pedagogical.


\section{Noises}


Consider two leads joined by a resistor. The value of its conductance, $G$, is to be meant, for the sake of precision, as corresponding to a two-terminal arrangement, meaning that the voltage $V$ is identified with the difference of chemical potentials between right movers on the left of the resistor and left movers on its right. We are interested in the average charge $\mean{Q}$ transported across the resistance in a time $T$, and in the variance $\cumul{Q^2} = \mean{Q^2}-\mean{Q}^2$, equivalently in the current $\mean{Q}/T$ and in the noise $\cumul{Q^2}/T$.

There are two types of noises:
\begin{enumerate}
 \item Equilibrium, or thermal, noise occurs in the absence of voltage, $V=0$, and at positive temperature $\beta^{-1}>0$. Then
\begin{equation}\label{jn}
 \mean{Q} = 0\,,\qquad\frac{\mean{Q^2}}{T} = \frac{2}{\beta}\,G\,.
\end{equation}
(Johnson \cite{Jo}, Nyquist \cite{Ny}). This is an early instance of the
fluctuation-dissipation theorem, those words being here represented as noise
and conductance. 

 \item Non-equilibrium, or shot, noise occurs in the reverse situation: $V\neq0$, $\beta^{-1} = 0$. Ohm's law states $\mean{Q} / T = G V$, while for the noise different expressions (corresponding to different situations) are available:
(a) classical shot noise
\begin{equation}
\label{ClassShot}
 \cumul{Q^2} = e\mean{Q}
\end{equation}
(Schottky \cite{Sch}), where $e$ is the charge of the carriers, say electrons. The result is interpreted on the basis of the Poisson distribution
\begin{equation*}
 p_n = \ep{-\lambda}\frac{\lambda^n}{n!}\,,\quad(n=0,1,2,\ldots)
\end{equation*}
of parameter $\lambda$, for which
\begin{equation*}
 \mean{n} = \lambda\,,\qquad\cumul{n^2}=\lambda\,.
\end{equation*}
Assuming that electrons arrive independently of one another, the number~$n$ of electrons collected in time~$T$ is so distributed, whence~(\ref{ClassShot}) for $Q = en$.

(b) quantum shot noise. Consider the leads and the resistor as modelled by a $1$-dimensional scattering problem with matrix
\begin{equation}\label{scmatrix}
 S = \left(\begin{array}{cc} r & t' \\ t & r' \end{array}\right)\,,
\end{equation}
where $r, t$ (resp. $r', t'$) are the reflection and transmission amplitudes from the left and from the right. Then
\begin{equation}
\label{QuantShot}
 \cumul{Q^2} = e \mean{Q}(1-|t|^2)
\end{equation}
(Khlus \cite{Kh}, Lesovik \cite{L}). In this case the result may be attributed to a binomial distribution with the success probability $p$ and with $N$ attempts:
\begin{align*}
 p_n &= \binom {N} {n} p^n(1-p)^{N-n}\,, \\
\mean{n} &= Np\,,\quad\cumul{n^2} = Np(1-p)\,.
\end{align*}
This yields~(\ref{QuantShot}) for $p = |t|^2$ being the probability of transmission. For small $p$ it reduces to~(\ref{ClassShot}). It should be noticed that in the case of thermal noise, the origin of fluctuations is in the source of electrons, or in the incoming flow, depending on the point of view. By contrast, in the interpretation of the quantum shot noise the flow is assumed ordered, as signified by the fixed number of attempts, and fluctuations arise only because of the uncertainty of transmission.
\end{enumerate}

We refer to~\cite{I} for a more complete exposition of these matters. We
conclude the section by recalling that noises are quantitative evidence to
atomism. Thermal noise determines $\beta^{-1} = k\cdot$temperature, and hence
Boltzmann's constant $k$ as well as Avogadro's number $N_0 = R/k$ (somewhat in
analogy to its determination from Brownian motion~\cite{Su, E}). Shot noise
determines the charge of carriers. In some instances of the fractional quantum
Hall effect this yielded $e/3$~\cite{Gl} or $e/5$ ~\cite{Rez}.


\section{A setup for counting statistics}\label{three}


Before engaging in quantum mechanical computations of the transported charge we should describe how it is measured, at least in the sense of a thought experiment. Consider a device (dot, resistor, or the like) connected to several leads, or reservoirs, one of which is distinguished (`the lead'). The measurement protocol consists of three steps:
\begin{itemize}
 \item measure the charge present {\it initially} in the lead, given a prepared state of the whole system.
 \item act on the system during some time by driving its controls (like gate voltages in a dot), but not by performing measurements. This includes the possibility of just waiting.
 \item measure the charge present in the lead {\it finally}.
\end{itemize}
The transported charge is then identified as the difference, $n$, of the outcome of the measurements. For simplicity we assume that $n$ takes only integer values, interpreted as the number of transferred electrons. Let $p_n$ be the corresponding probabilities. They are conveniently encoded in the generating function
\begin{equation}
\label{GenFunc}
\chi(\lambda) = \sum_{n\in\mathbb{Z}} p_n\ep{\iu\lambda n}
\end{equation}
of the moments of the distribution,
\begin{equation*}
\mean{n^k} = \left . \left(-\iu\frac{d}{d\lambda}\right)^k\chi(\lambda)\right|_{\lambda = 0}\,.
\end{equation*}
Similarly, $\log\chi(\lambda)$ generates the cumulants $\cumul{n^k}$, inductively defined by
\begin{equation*}
\mean{n^k} = \sum_{\mathcal{P}}\prod_{\alpha\in\mathcal{P}} \cumul{n^{|\alpha|}}\,,
\end{equation*}
where $\mathcal{P} = \{\alpha_1, \ldots,\alpha_m\}$ runs over all partitions of $\{1,\ldots,k\}$. Alternative protocols with measurements extending over time will be discussed later.


\section{Quantum description}\label{four}


The three steps of the procedure just described can easily be implemented quantum mechanically by means of two projective measurements and by a Hamiltonian evolution in between.

Let $\hilb$ be the Hilbert space of pure states of a system, $\rho$ a density matrix representing a mixed state, and $A = \sum_i \alpha_i P_i$ an observable with its spectral decomposition. A single measurement of $A$ is associated, at least practically, with the `collapse of the wave function' resulting in the replacement
\begin{equation}
\label{Collapse}
\rho\:\leadsto\: \sum_i P_i\rho P_i\,,
\end{equation}
where $\tr(P_i\rho P_i) = \tr (\rho P_i)$ is the probability for the outcome
$\alpha_i$. Two measurements of $A$, separated by an evolution given as a
unitary $U$, result in the replacement \cite{Schw}
\begin{equation}\label{repl}
\rho\:\leadsto\: \sum_{i,j} P_j U P_i\rho P_i U\str P_j \,,
\end{equation}
where $\tr(P_j U P_i\rho P_i U\str P_j) = \tr( U\str P_j U P_i\rho P_i)$ is the probability of the history $(\alpha_i, \alpha_j)$ of outcomes. We can so compute the moment generating function~(\ref{GenFunc}):
\begin{equation}
\label{GenFuncBis}
 \chi(\lambda) = \sum_{i,j}\tr( U\str P_j U P_i\rho
P_i)\ep{\iu\lambda(\alpha_j - \alpha_i)} = \sum_{i}\tr( U\str \ep{\iu\lambda A} U P_i\rho P_i)\ep{-\iu\lambda\alpha_i}\,.
\end{equation}
The expression simplifies if
\begin{equation}
\label{StateCommutes}
[A,\rho]=0\,;
\end{equation}
then $P_i \rho P_i = P_i\rho$, whence the r.h.s. of~(\ref{Collapse}) still equals $\rho$ (no collapse at first measurement) and 
\begin{equation}
\label{GenFuncSimpl}
\chi(\lambda) = \tr( U\str \ep{\iu\lambda A} U \ep{-\iu\lambda A}\rho)\,.
\end{equation}
If $\rho$ is a pure state, $\rho=\Omega(\Omega,\cdot)$, then
\begin{equation}
\label{GenFuncPure}
\chi(\lambda) = 
(\Omega,  U\str \ep{\iu\lambda A} U \ep{-\iu\lambda A}\Omega)\,.
\end{equation}
%

\section{Independent, uncorrelated fermions}\label{five}


We intend to apply~(\ref{GenFuncSimpl}) to many-body systems consisting of
fermionic particles which are uncorrelated in the initial state. The particles
shall contribute additively to the observable to be considered and evolve
independently of one another. The ingredients can therefore be specified at
the level of a single particle. At the risk of confusion we denote them like 
the related objects in the previous section: A Hilbert space $\hilb$ with
operators $A, U, \rho$. However, the meaning of $\rho$ is now that of a
$1$-particle density matrix $0\leq \rho\leq 1$ specifying an uncorrelated
many-particle state, to the extent permitted by the Pauli principle: any 
eigenstate of $|\nu\rangle$ of $\rho$, $\rho|\nu\rangle = \nu|\nu\rangle$, is
occupied in the many-particle state with probability given by its eigenvalue
$\nu$. Common examples are the vacuum $\rho = 0$ and, in terms of a
single-particle Hamiltonian $H$, the Fermi-Dirac distribution $\rho =
(1+\ep{\beta H})^{-1}$ or its zero temperature limit, $\beta^{-1}\to 0$, the
Fermi sea $\rho = \Theta(-H)$.

The corresponding many-particle objects are obtained through second quantization, which amounts to the following replacements:
\begin{align}
\label{MBhilb}
\hilb\:&\leadsto\:\mathcal{F}(\hilb) = \bigoplus_{n=0}^{\infty} \bigwedge^n\hilb \qquad \text{(Fock space)} \\
\label{MBsaOp}
A\:&\leadsto\:\dd\Gamma(A) \\
\label{MBunitary}
U\:&\leadsto\:\Gamma(U)
\end{align}
where $\dd\Gamma(A)$ and $\Gamma(U)$ act on the subspaces $\bigwedge^n\hilb \subset \mathcal{F}(\hilb)$ as
\begin{align*}
\dd\Gamma(A) &= \sum_{i=1}^n 1\otimes\cdots\otimes A \otimes\cdots \otimes 1\,, \\
\Gamma(U) &= U\otimes\cdots\otimes U\,.
\end{align*}
Moreover, the state is replaced as 
\begin{equation}
\label{MBstate}
\rho\:\leadsto\:\frac{\Gamma(\rho/\rho')}{\Tr_{\mathcal{F}(\hilb)}\Gamma(\rho/\rho')}\,,
\qquad (\rho' = 1-\rho)\,.
\end{equation}
Indeed, if $\rho$ splits with respect to $\hilb = \hilb_1\oplus\hilb_2$, then the many-body state~(\ref{MBstate}) factorizes w.r.t. $\mathcal{F}(\hilb) = \mathcal{F}(\hilb_1)\otimes \mathcal{F}(\hilb_2)$. In particular if $\rho|\nu\rangle = \nu|\nu\rangle$, this entails the following state on $\mathcal{F}[|\nu\rangle] = \oplus_{n=0}^1\wedge^n[|\nu\rangle]$
\begin{equation*}
\frac{1_0 + \frac{\nu}{\nu'}1_1}{1+\frac{\nu}{\nu'}} = \nu' 1_0 + \nu 1_1\,,
\end{equation*}
confirming that $\nu$ is the occupation number of $|\nu\rangle$. We note that
\begin{equation*}
\Tr_{\mathcal{F}(\hilb)}\Gamma(M) = \det{}_{\hilb}(1+M)\,,
\end{equation*}
provided that $M$ is a trace-class operator on $\hilb$, in which case the
r.h.s. is a Fredholm determinant. We will comment on that condition
later. Under the replacements~(\ref{MBhilb}-\ref{MBstate}) the assumption
$[A,\rho] = 0$ is inherited by the corresponding second quantized
observables, $[\dd\Gamma(A), \Gamma(\rho / \rho')] = 0$. As a result 
(\ref{GenFuncSimpl}) applies and becomes the Levitov-Lesovik formula
\begin{equation}
\label{LesLev}
\chi(\lambda) = \det (\rho' + \ep{\iu\lambda U\str A U}\ep{-\iu\lambda A}\rho)\,.
\end{equation}
Indeed,
\begin{align*}
\chi(\lambda) &= \frac{ \Tr_{\mathcal{F}(\hilb)}( \Gamma(U)\str \ep{\iu\lambda \dd\Gamma(A)} \Gamma(U) \ep{-\iu\lambda \dd\Gamma(A)} \Gamma(\rho / \rho') ) }{\Tr_{\mathcal{F}(\hilb)}\Gamma(\rho / \rho')} \\
 &=\frac{ \Tr_{\mathcal{F}(\hilb)}\Gamma(U\str \ep{\iu\lambda A} U \ep{-\iu\lambda A}\rho/\rho') }{\Tr_{\mathcal{F}(\hilb)}\Gamma(\rho / \rho')} = \frac{ \det(1+U\str \ep{\iu\lambda A} U \ep{-\iu\lambda A}(\rho/\rho')) }{ \det(1+(\rho/\rho')) } \\
 &= \det(\rho'+U\str \ep{\iu\lambda A} U \ep{-\iu\lambda A}\rho)\,.
\end{align*}

Before discussing the mathematical fine points of~(\ref{LesLev}), let us
compute the first two cumulants of charge transport. In line with the discussion in the previous section, let $A=Q$ be the projection onto single-particle states located in the distinguished lead. Then~(\ref{LesLev}) yields
\begin{align}
\mean{Q} &= -\iu\chi'(0) = \tr\rho(\Delta Q)\,, \nonumber \\
\cumul{Q^2} &= -(\log \chi)''(0) = \tr \rho(\Delta Q)(1-\rho)\Delta Q \nonumber \\
\label{2cumul}
&=\tr(\rho(1-\rho)(\Delta Q)^2) + \frac{1}{2}\tr(\iu[\Delta Q, \rho ] )^2\,,
\end{align}
where $\Delta Q = U\str Q U - Q$ is the operator of transmitted charge. The
split~(\ref{2cumul}) of the noise $\cumul{Q^2}$ into two separately non
negative contributions is of some interest (\cite{Buett} by a different 
approach,
\cite{Avronetal1}): The commutator $[\Delta Q, \rho]$ expresses the 
uncertainty of
transmission $\Delta Q$ in the given state $\rho$; the second term in 
(\ref{2cumul}) may thus be viewed as {\it shot noise}. The factor $\rho(1-\rho)$
expresses the fluctuation $\nu(1-\nu)$ in the occupation of single particle
states $|\nu\rangle$. It refers to the initial state, or source, and its term
may be viewed as {\it thermal noise}; indeed it vanishes for pure states,
$\rho=\rho^2$, while for $\rho = (1 + \ep{\beta H})^{-1}$ the energy width of
$\rho(1-\rho)$ is proportional to $\beta^{-1}$, \cf (\ref{jn}).


\section{Alternative approaches}


We present alternatives and variants of the two-step measurement procedures
discussed in Sect.~\ref{three}. We discuss them in the first quantized setting
of Sect.~\ref{four}. The corresponding second quantized versions can then
easily obtained from the replacements (\ref{MBhilb}-\ref{MBstate}).\\

i) \cite{LL1}
One could envisage a single measurement of the difference $U\str AU-A$. 
On the basis of (\ref{repl}) its generating function is
\begin{equation*}
\chi(\lambda) = \tr(\ep{\iu\lambda(U\str AU-A)}\rho)\,.
\end{equation*}
It remains unclear how to realize a von Neumann measurement for this
observable, since its two pieces are associated with two different times.
Moreover, its second quantized version 
\begin{equation*}
\chi(\lambda) =
\det (\rho' + \ep{\iu\lambda(U\str AU-A)}\rho)
\end{equation*}
generates cumulants which, as a rule beginning with $n=3$, differ from those
of (\ref{LesLev}).\\

ii) \cite{SchRa}
We keep the two-measurement setup, but refrain from making assumption 
(\ref{StateCommutes}), \ie, the first measurement is allowed to induce a
``collapse of the wave function''. We do however assume that the eigenvalues
$\alpha_i$ of $A$ are integers, in line with the application made at the 
end of the previous section, where $A\leadsto \dd\Gamma(Q)$ with $Q$ a 
projection. Then (\ref{GenFuncBis}) yields
\begin{align*}
 \chi(\lambda) &= \sum_{n,m}\tr( U\str \ep{\iu\lambda A} U P_n\rho P_m)
\delta_{mn}\ep{-\iu\lambda n}\\
&=\frac{1}{2\pi}\int_0^{2\pi}\dd\tau\,\tr( U\str \ep{\iu\lambda A} U
\ep{-\iu(\lambda+\tau)A}\rho\ep{\iu\tau A})
\end{align*}
by using $\delta_{mn}=(2\pi)^{-1}\int_0^{2\pi}\dd\tau\,\ep{\iu\tau(m-n)}$.\\

iii) \cite{LLL} 
Here neither (\ref{GenFunc}) nor (\ref{StateCommutes}) is assumed. The 
system is coupled to a
spin-\hbox{$\frac{1}{2}$} resulting in a total state space
$\hilb\otimes\mathbb{C}^2$. Specifically, the total Hamiltonian is obtained by
conjugating the system Hamiltonian by 
$\ep{-\iu\frac{\lambda}{2}A\otimes\sigma_3}$, where $\lambda$ is a coupling
constant and $\sigma_3$ a Pauli matrix; equivalently, the same holds true for
the evolution $U$, which becomes
\begin{equation*}
\widehat{U}=\ep{-\iu\frac{\lambda}{2}A\otimes\sigma_3}(U\otimes \Id)
\ep{\iu\frac{\lambda}{2}A\otimes\sigma_3}\,.
\end{equation*}
We note that 
\begin{equation*}
\widehat{U}(\psi\otimes|\sigma\rangle)=
(U_{\sigma\cdot\lambda}\psi)\otimes|\sigma\rangle\,,\qquad(\sigma=\pm 1)\,,
\end{equation*}
where $\sigma_3|\sigma\rangle=\sigma|\sigma\rangle$ and 
$U_\lambda=\ep{-\iu\frac{\lambda}{2}A}U\ep{\iu\frac{\lambda}{2}A}$.
The joint initial state is assumed of the form $\rho\otimes\rho_\mathrm{i}$
with $\rho$ being that of the system and 
\begin{equation*}
\rho_\mathrm{i}=
\bigl(\langle\sigma|\rho_\mathrm{i}|\sigma'\rangle\bigr)_{\sigma,\sigma'=\pm
1}
=\begin{pmatrix}\rho_{++}&\rho_{+-}\\\rho_{-+}&\rho_{--}\end{pmatrix}
\end{equation*}
that of the spin. The final state is 
$\widehat{U}(\rho\otimes\rho_\mathrm{i})\widehat{U}\str$ and, after tracing
out the system, 
\begin{equation*}
\rho_\mathrm{f}=\tr_\hilb\widehat{U}(\rho\otimes\rho_\mathrm{i})\widehat{U}\str
\end{equation*}
with matrix elements
\begin{equation*}
\langle\sigma|\rho_\mathrm{f}|\sigma'\rangle=
\tr(U_{\sigma\lambda}\rho U_{\sigma'\lambda}\str)
\langle\sigma|\rho_\mathrm{i}|\sigma'\rangle\,.
\end{equation*}
In other words,
\begin{equation*}
\rho_\mathrm{f}=
\begin{pmatrix}\rho_{++}&\rho_{+-}\chi(\lambda)\\
\rho_{-+}\chi(-\lambda)&\rho_{--}\end{pmatrix}
\end{equation*}
with 
\begin{equation*}
\chi(\lambda)=\tr(\ep{\iu\frac{\lambda}{2} A}  U\str 
\ep{-\iu\lambda A} U \ep{\iu\frac{\lambda}{2} A}\rho)\,.
\end{equation*}
We remark that $\chi(\lambda)$ agrees with (\ref{GenFuncSimpl}) under the
assumption (\ref{StateCommutes}) of the latter. It can be determined from the
average spin precession, since $\langle\sigma|\rho_\mathrm{f}|\sigma'\rangle$
reflects that measurement. On the other hand no
probability interpretation, \cf (\ref{GenFunc}), is available for
$\chi(\lambda)$, since its Fourier transform is non-positive in general
\cite{Na}.


\section{The thermodynamic limit}


The derivation of (\ref{LesLev}) was heuristic. It therefore seems 
appropriate to 
investigate whether the resulting determinant, cast as $\det(1+M)$, is 
well-defined, which is the case if $M$ is a trace-class operator. This happens
to be the case if the leads are of finite extent and the energy range finite, 
essentially because the single-particle Hilbert space becomes finite
dimensional. While these conditions may be regarded as effectively met in 
practice, it is
nevertheless useful to idealize these quantities as being infinite. There are
two physical reasons for that. First, any bound on these quantities ought 
to be irrelevant, because the transport occurs across the dot (compact in
space) and near the Fermi energy (compact in energy); second, the infinite
settings allows to conveniently formulate non-equilibrium stationary states.
However this idealization needs some care. In fact, in the attempt of extending
eq. (\ref{LesLev}) to infinite systems, the determinant becomes ambiguous and
ill-defined. The cure is a regularization which rests on the heuristic
identity 
\begin{equation}
\tr(U\str\rho QU-\rho Q)=0\,,
\label{fp1}
\end{equation}
obtained by splitting the trace and using its cyclicity. It consists in 
multiplying the determinant by 
\begin{equation}
\det(\ep{-\iu\lambda U\str\rho QU})\cdot\det(\ep{\iu\lambda \rho Q})=
\ep{-\iu\lambda\tr(U\str\rho QU-\rho Q)}=1\,,
\label{fp2}
\end{equation}
thereby placing one factor on each of its sides. The straightforward result is 
(\cite{Avronetal2}, and in the zero-temperature case \cite{MuzykantskyAdamov})
\begin{equation}
\label{regdet}
\chi(\lambda)=\det(\ep{-\iu\lambda \rho_U Q_U}\rho'\ep{\iu\lambda \rho
Q}+\ep{\iu\lambda \rho'_U Q_U}\rho\ep{-\iu\lambda \rho' Q})\,,
\end{equation}
where $\rho'=1-\rho$, $\rho_U=U\str\rho U$, and similarly for $\rho'_U$ and
$Q_U$.\\

\noindent
{\bf Remarks.} 1. We observe a manifest particle-hole symmetry: 
\begin{equation*}
\chi_\rho(\lambda)=\chi_{\rho'}(-\lambda)\,.
\end{equation*}
2. We will see that the determinant (\ref{regdet}) is Fredholm under
reasonable hypotheses. \\
3. The regularization bears some resemblance to $\det_2(1+M)
=\det(1+M)\ep{-\tr M}$, though the latter typically changes the value of the
determinant.\\

To the extent that the regularization is regarded as a modification at all, it
affects only the first cumulant, because the term 
$-\iu\lambda\tr(\rho_U Q_U-\rho Q)$, which by (\ref{fp2}) has been added to 
the generating function $\log \chi(\lambda)$, is linear in $\lambda$. 
The mean is thus
changed from $\langle n\rangle=\tr\rho(Q_U-Q)$ to 
$\langle n\rangle=\tr (\rho-\rho_U)Q_U$. In line with Sections~\ref{three} and 
\ref{five} we interpret $Q$ as the projection onto single-particle states
in the distinguished lead and $U$ as the evolution preserving the initial
state $\rho$, except for changes in the dot. We then expect that $Q_U-Q$
is non-trivial on states of any energy, while $\rho-\rho_U$ is so only on 
states 
which are located near the dot and near the Fermi energy. As a result, the
second expression for $\langle n\rangle$, but not the first one, appears to be
well-defined.


\section{A more basic approach}

The regularization (\ref{fp1}) remains an ad hoc procedure, though it may be
motivated as a cancellation between right and left movers, 
see~\cite{Avronetal2}.
The point we wish to make here is that eq. (\ref{regdet}) is obtained without
any recourse to regularization if the second quantization is based upon a
state of positive density (rather than the vacuum, \cf Sect.~\ref{five}), 
as it is appropriate for an open system. 

To this end let us briefly recall the defining elements of quantum 
mechanics of infinitely many degrees of freedom: (local) observables
are represented by elements of a C*-algebra $\mathcal{A}$ and states by  
normalized, positive, continuous linear functionals on $\mathcal{A}$. 
A state $\omega$, together with its local perturbations, may be given a 
Hilbert space realization through the GNS construction: it consists of a
Hilbert space $\hilb_\omega$, a representation $\pi_\omega$ of $\mathcal{A}$
on $\hilb_\omega$, and a cyclic vector $\Omega_\omega\in\hilb_\omega$ such
that
\begin{equation*}
\omega(A)=(\Omega_\omega, \pi_\omega(A)\Omega_\omega)\,,\qquad (A\in\mathcal{A})\,.
\end{equation*}
Notice that the state $\omega$ is realized as a vector, $\Omega_\omega$, 
regardless of whether it is
pure. Rather, it is pure iff the commutant 
$\pi_\omega(\mathcal{A})'\subset\mathcal{L}(\hilb_\omega)$ is trivial. 
The closure of $\pi_\omega(\mathcal{A})$ yields the von Neumann algebra 
$\overline{\pi_\omega(\mathcal{A})}$. Besides of local observables 
$\pi_\omega(A)$ it also contains 
some global ones, whose existence and meaning presupposes $\omega$. An 
example occurring in the following is the charge present in the (infinite)
lead in excess of the (infinite) charge attributed to $\omega$.

The C*-algebra of the problem at hand is $\alg{\hilb}$, the algebra of 
canonical anti-commutation relations over the single-particle 
Hilbert space $\hilb$. It is the algebra with unity generated by the elements 
$a(f)$, $a(f)\str$ (anti-linear, resp. linear in $f\in\hilb$) satisfying 
\begin{equation*}
\{a(f),a\str(g)\}=(f,g)\Id\,,\qquad \{a(f),a(g)\}=0=\{a\str(f),a\str(g)\}\,.
\end{equation*}
A unitary $U$ induces a *-automorphism of the algebra by $a(f)\mapsto a(Uf)$
(Bogoliubov automorphism).
A single-particle density matrix $0\le\rho\le \Id$ defines a state 
$\omega$ on $\alg{\hilb}$ through 
\begin{equation*}
\omega(a\str(f)a(g))=(g, \rho f)\,,\qquad
\omega(a(f)a(g))=0=\omega(a\str(f)a\str(g))
\end{equation*}
and the use of Wick's lemma for the ccomputation of higher order correlators. 
States of this form are known as 
gauge-invariant quasi-free states; they describe uncorrelated fermions.
It is possible to give an explicit construction of their GNS representation,
known as Araki-Wyss representation, but we will not need it. 

For clarity we formulate the main result first for pure state and then for 
mixed states. In both cases we assume 
$[\rho,Q]=0$, \cf (\ref{StateCommutes}).\\

\noindent
{\bf Theorem} (Pure states). Let $\rho=\rho^2$. We assume that 
\begin{equation}\label{trcl}
\rho-U\rho U\str
\end{equation}
is trace class. Then
\begin{enumerate}
\item The 
algebra automorphisms $a(f)\mapsto a(Uf)$ and
$a(f)\mapsto a(\ep{\iu\lambda Q}f)$ are unitarily implementable: There exists
(non-unique) unitaries $\widehat{U}$ and $\ep{\iu\lambda \widehat{Q}}$ on
$\mathcal{H}_\omega$ such that
\begin{equation*}
\widehat{U}\pi_\omega(a(f))=\pi_\omega(a(Uf))\widehat{U}\,,\qquad
\ep{\iu\lambda \widehat{Q}}\pi_\omega(a(f))=
\pi_\omega(a(\ep{\iu\lambda Q}f))\ep{\iu\lambda \widehat{Q}}\,.
\end{equation*}
\item $\widehat{Q}$ is an observable, in the sense that any bounded function
thereof is in $\overline{\pi_\omega(\alg{\hilb})}$.
\item \label{iii} The above properties define $\widehat{U}$ uniquely up to a
phase and $\widehat{Q}$ up to an additive
constant.  
\item The generating function of cumulants, \cf (\ref{GenFuncPure}), equals the
regularized determinant (\ref{regdet}):
\begin{equation*}
(\Omega_\omega,\widehat{U}\str\ep{\iu\lambda \widehat{Q}}\widehat{U} \ep{-\iu\lambda \widehat{Q}}\Omega_\omega)=\det(\ep{-\iu\lambda \rho_U Q_U}\rho'\ep{\iu\lambda \rho
Q}+\ep{\iu\lambda \rho'_U Q_U}\rho\ep{-\iu\lambda \rho' Q})\,,
\end{equation*}
where the determinant is Fredholm.
\end{enumerate}

Eq.~(\ref{trcl}) demands that the evolution $U$ preserves $\rho$,
except for creating excitations of finite energy within an essentially finite 
region of space. This assumption is appropriate for the evolution 
induced by a compact device operating smoothly during a finite time interval.

The generalization to mixed states is as follows.\\

\noindent
{\bf Theorem} (Mixed states). Let $0<\rho<1$. Assume, instead of (\ref{trcl}),
that $\rho^{1/2}-U\rho^{1/2} U\str$ and $(\rho')^{1/2}-U(\rho')^{1/2} U\str$
are trace class; moreover that 
\begin{equation}
\label{trcl1}
(\rho\rho')^{1/2}Q
\end{equation}
is, too. Then the above results (i-iv) hold true, upon replacing (iii) by 
\begin{itemize}
\item
Properties (i-ii) define $\widehat{U}$ uniquely up to left 
multiplication with an element from the commutant
$\pi_\omega\bigl(\alg{\hilb}\bigr)'$, and $\widehat{Q}$ up to an additive
constant. In particular, $\widehat{U}\str \ep{\iu\lambda
\widehat{Q}}\widehat{U} \ep{-\iu\lambda \widehat{Q}}$ is unaffected
by the ambiguities. 
\end{itemize}

Notice that the most general case, $0\le\rho\le 1$, is not covered.
The physical origin of the extra assumption~(\ref{trcl1})
needed in the mixed state case is as follows. In both cases, pure or mixed, 
the expected charge contained in a portion of the lead is of order of
its length $L$, or zero if renormalized by subtraction of a
background charge. In the pure case however, the Fermi sea is an
eigenvector of the charge operator, while for the mixed state, the
variance of the charge must itself be of order $L$, because the
occupation of the one-particle states is fluctuating, due to
$\rho\rho'\neq0$. Hence, in this latter situation, the measurement of the
renormalized charge yields finite values only as long as $L$ is
finite, of which eq.~(\ref{trcl1}) is a mathematical
abstraction. 
In the limit $L\to\infty$ all but a finite part of the fluctuation of the
source is affecting the transmitted noise. That suggests perhaps that there is 
a better formulation of the result. Indeed, the expression for the 
transmitted noise, \cf the first term (\ref{2cumul}), is finite if
$(U\str QU-Q)(\rho\rho')^{1/2}$ is trace class. This condition turns out to be
sufficient for property (i), for making the determinant Fredholm and 
$\widehat{U\str QU-Q}$ an observable, but not for (ii, iii). 

For proofs we refer to \cite{Avronetal2}.

\section{An application}

We discuss a very simple application to illustrate the working of the
regularization. The system consists of two leads in guise of circles of length
$T$, joined at one point. Particles run in the positive sense along the
circles $C$ at velocity $1$, whence it takes them time $T$ to make a turn, 
and may scatter from one to the other circle at the junction. Initially states
in the two circles are populated up to Fermi energies $\mu_L<\mu_R$. This 
is formalized as follows. The single particle Hilbert space is 
\begin{equation*}
\hilb=L^2(C)\oplus L^2(C)\ni\psi=\begin{pmatrix}\psi_L(x)\\ 
\psi_R(x)\end{pmatrix}\,,
\end{equation*}
the evolution over time $T$ is 
\begin{equation*}
(U\psi)(x)=S\psi(x)
\end{equation*}
with $S$ as in (\ref{scmatrix}). The momentum operator is $p=-\iu d/dx$ and 
the initial state is $\rho=\rho_L\oplus \rho_R$ with 
$\rho_i=\theta(\mu_i-p)$, ($i=L,R$). The projection onto the right lead is
$Q=0\oplus 1$. 

Quite generally, for $\rho=\rho^2$ a pure state, eq. (\ref{regdet}) reads
\begin{equation*}
\chi(\lambda)=\mydet_\hilb(1+(\ep{-\iu\lambda}-1)Q_U \rho_U\rho'+
(\ep{\iu\lambda}-1)Q_U \rho'_U \rho)\,,
\end{equation*}
and in the present situation that determinant reduces to
\begin{equation*}
\chi(\lambda)=\mydet_{L^2(C)}(1+(\ep{\iu\lambda}-1)\rho_R'
\rho_L|t|^2)\,.
\end{equation*}
It is to be noted that $\rho_R'\rho_L$ selects a finite energy interval, 
$(\mu_L,\mu_R]$, unlike the determinant without regularization. 
Using eigenstates of momentum $p\in(2\pi/T)\mathbb{Z}$ we find 
\begin{align*}
\chi(\lambda)&=\prod_{\mu_L<p\le \mu_R}(1+(\ep{\iu\lambda}-1)|t|^2)\\
&=(1-|t|^2+\ep{\iu\lambda}|t|^2)^N=(q+\ep{\iu\lambda}p)^N
\end{align*}
with $N=\#\{p\mid\mu_L<p\le \mu_R\}\cong (\mu_L-\mu_R)T/2\pi$.
This is a binomial distribution with probability $p=|t|^2$ and $N$ attempts,
reproducing (\ref{QuantShot}). In particular, it yields Ohm's law 
$\mean{Q} / T = G V$ with $G=(2\pi)^{-1}|t|^2$ \cite{La, ES}.

\end{document}